\documentclass[]{IEEEtran}\pdfoutput=1

\usepackage[utf8]{inputenc}
\usepackage[T1]{fontenc}
\usepackage{tabularx}
\usepackage{framed}
\usepackage{subfigure}
\usepackage{color}
\usepackage{xspace}
\usepackage{booktabs}
\usepackage{multirow}
\usepackage{paralist}
\usepackage[colorlinks,linkcolor=blue,citecolor=blue,urlcolor=blue]{hyperref}

\usepackage{listings}
\usepackage{textcomp} 
\usepackage{textpos}
\usepackage{comment}
\usepackage[usenames,dvipsnames]{xcolor}

\definecolor{orange}{RGB}{255,127,0}
\definecolor{grey}{RGB}{135,135,135}

\usepackage[overload]{empheq}

\newtheorem{defn}{Definition}[section]

\usepackage{letltxmacro}
\newcommand*{\SavedLstInline}{}
\LetLtxMacro\SavedLstInline\lstinline
\DeclareRobustCommand*{\lstinline}{
  \ifmmode
    \let\SavedBGroup\bgroup
    \def\bgroup{
      \let\bgroup\SavedBGroup
      \hbox\bgroup
    }
  \fi
  \SavedLstInline
}

\definecolor{orange}{RGB}{255,127,0}
\definecolor{grey}{RGB}{135,135,135}

\newcommand{\thetitle}
  {Casper:  Debugging Null Dereferences with Dynamic Causality Traces}

\newcommand\ourtool{{\sc Casper}\xspace}
\newcommand\projecturl{\url{https://github.com/Spirals-Team/casper}}

\newcommand\nullliteral{L\xspace}
\newcommand\nullassign{A\xspace}
\newcommand\nullparamentry{$P_e$\xspace}
\newcommand\nullparamcall{$P_i$\xspace}
\newcommand\nullreturn{R\xspace}
\newcommand\nullunboxed{U\xspace}
\newcommand\nullderef{\textbf{D}\xspace}

\newcommand\exorcise{X\xspace}

\newcommand\nullv{\texttt{null}\xspace}
\newcommand\nullvs{\texttt{null}s\xspace}

\newcommand\nbtransfo{11\xspace}

\title{\thetitle}

\author{
Benoit Cornu, Earl T. Barr\dag, Lionel Seinturier*, Martin Monperrus*\\ 
*University of Lille \& INRIA\\
\dag University College London\\
contact:martin.monperrus@univ-lille1.fr
}

\begin{document}

\maketitle

\textbf{Abstract:} Fixing a software error requires understanding its root cause. In this paper, we introduce ``causality traces'', crafted execution traces augmented with the information needed to reconstruct the causal chain from the root cause of a bug to an execution error. We propose an approach and a tool, called \ourtool, based on code transformation, which dynamically constructs causality traces for null dereference errors. The core idea of \ourtool is to replace \lstinline+null+s with special values, called “ghosts”, that track the propagation of the nulls from inception to their error-triggering dereference.  Causality traces are extracted from these ghosts. We evaluate our contribution by providing and assessing the causality traces of 14 real null dereference bugs collected over six large, popular open-source projects. Over this data set, \ourtool builds a causality trace in less than 1 second.

\section{Introduction}

In isolation, software errors are often annoyances, perhaps costing one person a few hours of work when their accounting application crashes.  Multiply this loss across millions of people; consider that even scientific progress can be delayed or derailed by software error~\cite{merali2010computational}:  in aggregate, these errors are costly to society as a whole.

Fixing these errors requires understanding their root cause, a process that we call causality analysis. Computers are mindless accountants of program execution, yet do not track the data needed for causality analysis. This work proposes to harness computers to this task.  We introduce \emph{``causality traces''}, execution traces augmented with the information needed to reconstruct a causal chain from a root cause to an execution error.  We construct causality traces over \emph{``ghosts''}, an abstract
data type that can replace a programming language's special values, like \nullv or \lstinline+NaN+. Ghosts replace such values and track operations applied to itself, thereby collecting a causality trace whose analysis reveals the root cause of a bug.  To demonstrate the feasibility and promise of causality traces, we have instantiated ghosts for providing developers with causality traces for null deference errors, they are ``null ghosts''. 

Anecdotally, we know that null dereferences are frequent runtime errors.
Li et al. substantiated this conventional wisdom, finding that 37.2\% of all memory errors in Mozilla and and Apache are null dereferences~\cite{li2006have}.
Kimura et al. \cite{Kimura2014returnnull} found that there are between one and four null checks per 100 lines of code on average.
A null dereference runtime error occurs when a program tries to read memory
using a field, parameter, or variable that
points to nothing --- ``\nullv'' or ``none'', depending on the language.
For example, on October 22 2009, a developer working on the Apache Common Math
open source project encountered an null pointer exception and reported it as bug \#305\footnote{\url{https://issues.apache.org/jira/browse/MATH-305}}.

In low-level, unmanaged runtime environments, like assembly or C/C++, null
dereferences result in a dirty crash, e.g. a segmentation fault.
In a high-level, managed runtime environment such as Java, .NET, etc., a null dereference triggers an exception.
Programs often crash when they fail to handle
null dereference exceptions properly~\cite{bond2007tracking}.

When debugging a null dereference, the usual point of departure is a stack trace that contains all the methods in the execution stack at the point of the dereference.
This stack trace is decorated with the line numbers where each method was called.
\autoref{fig:math-369-st} gives an
example of such a stack trace and shows that the null dereference happens at
line 88 of \lstinline+BisectionSolver+.

\begin{lstlisting}[captionpos=b, numbers=left, breaklines=true, basicstyle=\scriptsize,
escapeinside={@}{@}, label={fig:math-369-st}, caption={The standard stack trace
of a real null dereference bug in Apache Commons Math}, float,
xleftmargin=5mm]
Exception in thread "main" java.lang.NullPointerException
	at [..].BisectionSolver.solve(88)
	at [..].BisectionSolver.solve(66)
	at ...
\end{lstlisting}

\begin{lstlisting}[
breaklines=true,escapechar=@, label={fig:math-369-st-2},
float, xleftmargin=5mm, basicstyle=\scriptsize,
caption={What we propose: a causality trace, an extended stack trace
  that contains the root cause.}]
Exception in thread "main" java.lang.NullPointerException 
@\textbf{For parameter : f}@ // symptom
	at [..].BisectionSolver.solve(88)
	at [..].BisectionSolver.solve(66)
	at ...
@\textbf{Parameter f bound to field UnivariateRealSolverImpl.f2}@
	@\textbf{at [..].BisectionSolver.solve(66)}@ 	
@\textbf{Field f2 set to null}@
	@\textbf{at [..].UnivariateRealSolverImpl.<init>(55)}@ // cause
\end{lstlisting}

Unfortunately, this stack trace only contains a partial snapshot of the program execution when the null dereference occurs, and not its root cause.
In \autoref{fig:math-369-st}, the stack trace says that a variable is null at
line 88, but not when and what assigned ``null'' to the variable.
Indeed, there may be a large gap between the symptom of a null dereference and its root cause \cite{bond2007tracking}.  In our evaluation, we present 7 cases where the patch for fixing the root cause of a null dereference error is not in any of the stack trace's method.  
This gap exactly is an instance of  Eisenstadt's cause/effect chasm~\cite{eisenstadt1997my} for a specific defect class: null dereference errors.

This ``null dereference cause/effect chasm'' has two dimensions.
The first is temporal: the symptom may happen an arbitrarily long time after its
root cause, e.g. the dereference may happen ten minutes and one million method
executions after the assignment to null. In this case, the stack trace is a
snapshot of the execution at the time of the symptom, not at the time of the root cause.
The second is spatial:
the location of the symptom may be arbitrarily far from the location of the root
cause. For example, the null dereference may be in package \lstinline{foo} and
class \lstinline{A} while the root cause may be in package \lstinline{bar} and
class \lstinline{B}.
The process of debugging null dereferences consists of tracing the link in space and time between the symptom and the root cause.

A \emph{causality trace} 
captures the complete history of the propagation of a \nullv value that is
incorrectly dereferenced.
\autoref{fig:math-369-st-2}
contains such a causality trace.
In comparison to \autoref{fig:math-369-st}, it contains three additional pieces of information.
First, it gives the exact name, here \lstinline+f+, and kind, here
parameter (local variable or field are other possibilities), of the null variable.
Second, it explains the origin of the parameter, the call to \lstinline{solve}
at line 66 with field \lstinline{f2} passed as parameter.
Third, it gives the root cause of the \nullv dereference:
the assignment of null to the field \lstinline{f2} at line 55 of class
\lstinline{UnivariateRealSolverImpl}.
Our causality traces contain several kinds of trace elements, of which
\autoref{fig:math-369-st-2} shows only three:
 the name of the wrongly dereferenced variable, the flow of nulls through parameter bindings, and null assignment.
\autoref{sec:contribution} details the rest.

In this paper, we present a novel tool, called \ourtool, to collect null causality traces.
The tool is going to be used at debugging time by developers.
It takes as input the program under debug and a main routine that triggers the null dereference.
It then outputs the causality trace of the null dereference.
It first instruments the program under debug by replacing \nullvs with ``ghosts''
that track causal information during execution.  We have named 
our tool \ourtool, since it injects ``friendly'' ghosts into buggy 
programs.
To instrument a program, \ourtool applies a set of \nbtransfo source code transformations tailored for building causal connections.
For instance, \lstinline+o = externalCall()+ is transformed into
\lstinline+o = NullDetector.check(externalCall())+, where the method
\lstinline{check} stores
causality elements in a null ghost (\autoref{sec:explain-nullghost}) and assigns it to
\lstinline{o} if \lstinline{externalCall} returns \nullv. \autoref{sec:transforms} details these tranformations.

We evaluate our contribution \ourtool by providing and assessing
the causality traces of 14 real null dereference bugs collected over six large, popular
open-source projects.  We collected these bugs from these project's bug reports, retaining those we were able to reproduce.
\ourtool constructs the complete causality trace for 13 of these 14 bugs.
For 11 out of these 13 bugs, the causality trace contains the location of the actual fix made by the developer.

Furthermore, we check that our \nbtransfo source code transformations do not change the semantics of the program relative to the program's 
test suite, by running the program against that test suite after transformation and confirming that it still passes. The limitations of our approach are discussed in \autoref{sec:imp} and its overhead in \autoref{sec:overhead}.

\medskip
To sum up, our contributions are:\vspace{-0.1cm}

\begin{itemize}\itemsep0em
\item The definition of causality traces for null dereference errors
\item A set of source code transformations designed and tailored for collecting the causality traces.
\item \ourtool, an implementation in Java of our technique.
\item An evaluation of our technique on real null dereference bugs collected over 6 large open-source projects.
\end{itemize}

\ourtool and our dataset can be downloaded from \projecturl.

\section{Casper's Null Debugging Approach}
\label{sec:contribution}

\ourtool tracks the propagation of \nullvs used during application execution in
a causality trace.  
A \emph{null dereference causality trace} is the sequence of
language constructs traversed during execution from the source of the \nullv to its erroneous dereference.  
By building this trace, \ourtool generalizes dynamic taint analysis to answer not only whether a null can reach a dereference, but \emph{how} a dereference is reached\footnote{We take the source of a null to be a given, and do not concern ourselves with its cause.  Under this assumption, Casper’s traces, which answer the question of how a dereference is reached, are causal.}
These traces speed the localization of the root cause of null dereferences errors. 

Our idea is to replace \nullvs with objects whose behavior, from
 the
application's point of view, is same as a \nullv, except that they store a
causality trace, defined in \autoref{sec:nullcausalitytrace}.  We called these
objects \emph{null ghosts} and detail them in
\autoref{sec:explain-nullghost}.  \ourtool rewrites the program under debug to
use null ghosts and to store a \nullv's causality trace in those null ghosts,
\autoref{sec:transforms}. An additional set of semantic preserving transformation are required (\autoref{sec:sp}).
Finally, we discuss \ourtool's realization in
\autoref{sec:imp}.  We instantiated \ourtool in Java and therefore tailored 
our presentation in this section to Java.

\subsection{Null Dereference Causality Trace}
\label{sec:nullcausalitytrace}

To debug a complex null dereference, the developer has to understand the
history of a guilty null from its creation to its problematic dereference.
She has to know the details of the \nullv's propagation, i.e. why and when each
variable became null at a particular location.  We call this history the ``null
causality trace'' of the null dereference.

\begin{defn}\label{def:ndctrace}
A \emph{null dereference causality trace} is the temporal sequence of
language constructs traversed by a dereferenced \nullv.
\end{defn}

\begin{table}[t]
\centering
\begin{tabularx}{\columnwidth}{ccl}
\toprule
Mnemonic & Description & Examples \\ 
\midrule
\multirow{4}{*}{\nullliteral} & \multirow{4}{*}{null literal} 
	& \lstinline+x = null;+ \\
 & & \lstinline+Object x; // Implicit null+ \\
 & & \lstinline+return null;+ \\
 & & \lstinline+foo(null);+ \\ 
 & & \\[-1ex]

\multirow{2}{*}{\nullparamentry} & \multirow{2}{*}{null at entry}
	& \lstinline+void foo(int x)+ \\
 & & \lstinline+\{+ \ldots \lstinline+\}+ \\
 & & \\[-1ex]

\multirow{2}{*}{\nullparamcall} & null 
  & \multirow{2}{*}{\lstinline+foo(+$e$\lstinline+)+} \\
 & at invocation & \\
 & & \\[-1ex]

\multirow{3}{*}{\nullreturn} & \multirow{3}{*}{null return}
	& \lstinline+x = foo() // foo returns null+ \\
 & & \lstinline+foo().bar()+ \\
 & & \lstinline+bar(foo())+ \\
 & & \\[-1ex]

\multirow{2}{*}{\nullunboxed} & \multirow{2}{*}{unboxed \nullv} 
	& \lstinline+Integer x =+ $e$; \\
 & & \lstinline+int y = x+ \\ 
 & & \\[-1ex]

\nullassign & null assignment & \lstinline+x =+\ $e$; \\
 & & \\[-1ex]

\multirow{2}{*}{\nullderef} & \textbf{null} & \lstinline+x.foo()+ \\ 
 & \textbf{dereference} &  \lstinline+x.field+ \\ 
 & & \\[-1ex]

\exorcise & \textbf{external call} & \lstinline+lib.foo(+$e$\lstinline+)+ \\
\bottomrule
\end{tabularx}
\caption{Null-propagating language constructs.
We use $e$ to denote an arbitrary expression. 
In all cases but  X, where $e$ appears, 
a \nullv propagates only if $e$ evaluates to  \nullv.  
\nullassign is generic assignment; it is the least specific \nullv-propagating 
language construct.  
}
\label{table:causalitylinks} 
\end{table}

Developers read and write source code. Thus, source code is the natural medium in
which developers reason about programs for debugging.  In
particular, a \nullv propagates through moves or copies, which are realized via
constructs like assignments and parameter binding.  This is why \ourtool defines causal links in a null
causality trace in terms of traversed language constructs.  \autoref{table:causalitylinks} depicts
language constructs through which \nullvs originate, propagate, and
trigger null pointer exceptions.  

  Therefore, \nullvs can originate in hard-coded
null literals (\nullliteral), and in external library
return (\nullparamcall) or callbacks (\nullparamentry).  In our causality
abstraction, these links are the root causes of a null dereference. Recall that Java forbids pointer arithmetic, so we do not consider this case.

A \nullv propagates through method parameters, returns (\nullreturn),
and unboxing (\nullunboxed).  With the least specificity, a \nullv can
propagate through source level assignment (\nullassign).
\nullderef denotes the erroneous dereference of a \nullv.

When \nullvs are passed as parameter, they can be detected at parameter binding bound at a call site (\nullparamcall) and method  entry (\nullparamentry).
The reasons are twofold.
First, we don't make any assumption on the presence, the observability and the manipulability of library code, so even in the presence of external libraries, we can trace that nulls are sent to them (\nullparamcall).
Second, it enables to decipher polymorphism in the traces, we trace the actual method that has been called.

Let us consider the snippet ``\lstinline+x = foo(bar())); ... x.field+'' and assume that
\lstinline+x.field+ throws an NPE.  The resulting null causality trace is
\nullreturn-\nullreturn-\nullassign-\nullderef (return return assignment dereference).
Here, the root cause is the return of the method \lstinline+bar+\footnote{The
root cause is not somewhere above the \lstinline+return+ in \lstinline+bar+'s
body or the causality trace would necessarily be longer.}. The trace of
\autoref{fig:math-369-st-2}, discussed in introduction, is
\nullliteral-\nullassign-\nullparamentry-\nullparamcall-\nullderef.  
\nullliteral and \nullassign are redundant in this case, since a single assignment statement directly assigns a \nullv to the field \lstinline+f2+; 
\nullparamentry and \nullparamcall are also redundant since no library call is
involved.  Post-processing removes such redundancy in a causality trace before \ourtool presents the trace to a user.

\ourtool decorates the \nullliteral, \nullassign, \nullparamcall, and
\nullunboxed ``links'' in a null dereference causality trace with the target
variable name and the signature of the expression assigned to the target
For each causal link, \ourtool also collects the location of the language constructs (file, line) as well as the name and the stack of the current thread.  Consequently, a causality trace contains a
temporally ordered set of information and not just the stack at the point in
time of the null dereference.  In other words, a causality trace contains a
\nullv's root cause and not only the stack trace of the symptom.

A causality trace is any chain of these causal links. A trace 
\begin{inparaenum}[\itshape a\upshape)]
\item starts with a \nullliteral or, in the presence of an external library, 
  with \nullreturn or \nullparamentry ;
\item may not end with a dereference (if the null pointer exception is caught, the null can continue to propagate); and
\item may contain a return, not preceded by a method parameter link, when a void
  external method returns \nullv.
\end{inparaenum}
A causality trace can be arbitrarily long.

\begin{table*}
\begin{tabularx}{\textwidth}{p{3.6cm}|X}
\textbf{Method} & \textbf{Explanation}  \\
\hline
nullAssign(x, position) & logs whether x is null at this assignment, returns x if x is a valid object or a ghost, or a new ghost if x is null\\
nullParam(x, position) & logs whether x is null when passed as parameter, returns x  if x is a valid object or a ghost, or a new ghost if x is null\\
nullPassed(x, position) & logs whether x is null when received as parameter at this position, returns void\\
nullReturn(x, position) & logs whether x is null when returned at this position, returns x  if x is a valid object or a ghost, or a new ghost if x is null\\
exorcise(x, position) & logs whether x is null when passed as parameter to a library call at this position, returns "null" if x is a ghost, or x\\ nullUnbox(x, position) & throws a null pointer exception enriched with a causality trace if a null ghost is unboxed \\
nullDeref(x, position) & throws a null pointer exception enriched with a causality trace if a field access is made on a null ghost \\
\end{tabularx}
  \caption{Explanations of the Methods Injected with Code Transformation.}
\label{tab:explanation-injected-methods}
\end{table*}

\begin{lstlisting}[float, numbers=none,captionpos=b,
caption={For each class of the program under debug, \ourtool generates a null ghost class to replace \nullvs.},
label=fig:example-ghost-class]
// original type
public MyClass{
    private Object o;
    public String sampleMethod(){
       ...
} }

// corresponding generated type
public MyGhostClass extends MyClass{
    public String sampleMethod(){
        // enriches the causality trace to log
        // that this null ghosts was dereferenced
        computeNullUsage();
        throw new CasperNullPointerException();
    }
    ... // for all methods incl. inherited ones
}
\end{lstlisting}

\subsection{Null Ghosts}
\label{sec:explain-nullghost}

The special value \nullv is the unique bottom element of Java's
nonprimitive type lattice.  Redefining \nullv to trace the propagation
of \nullvs during a program's execution, while elegant, is infeasible,
since it would require the definition and implementation of a new language, with
all the deployment and breakage of legacy applications that entails.  

For sake of applicability, we leave our host language, here Java, untouched and we use un vanilla unmodified Java virtual machine.
We use rewriting to create a
\emph{null ghost} to ``haunt'' each class defined in a codebase. A null ghost is
an object that 1) contains a null causality trace and 2) has the same observable
behavior as a null value.  To this end, a ghost class contains a queue and
overrides all methods of the class it haunts to throw null pointer exceptions.

Listing \ref{fig:example-ghost-class} illustrates this idea.  \ourtool creates the ghost class 
\lstinline+MyGhostClass+ that extends the application class MyClass.  All
methods defined in the application type are overridden in the new type (e.g.,
\texttt{sampleMethod}) as well as all other methods from the old class (See \autoref{sec:imp}).  The new methods
completely replace the normal behavior and have the same new behavior.  First,
the call to \texttt{computeNullUsage} enriches the causality trace with a causal
element of type \nullderef by stating that this null ghosts was dereferenced.
Then, it acts as if one has dereferenced a null value: it throws a
\lstinline+CasperNullPointerException+ (a special version of  the Java error
thrown when one calls a method on a null value called
\lstinline+NullPointerException+), which is discussed next.  
Also, a null ghost is an
instance of the marker interface NullGhost. This marker interface will be used
later to keep the same execution semantics between real null and null ghosts.

\subsection{{\large \ourtool}'s Transformations}
\label{sec:transforms}

\ourtool's transformations instrument the program under debug to 
detect \nullvs and construct null dereference causality traces dynamically, while preserving
its semantics.

\begin{figure*}[t]
\lstset{basicstyle=\ttfamily\scriptsize} 
\begin{subequations}
  \begin{align}[left = {T(e) = \empheqlbrace}]
  & e_1\ \lstinline+== null+\ \lstinline+||+ ~e_1\ \lstinline+instanceof NullGhost+ 
  && \quad\text{if } e = e_1\ \lstinline+== null+
    \label{eq:tr:equalnull} \\
  & e_1\ \lstinline+instanceof MyClass+\ \mathtt{\&\&}~!(e_1\ \lstinline+instanceof NullGhost+) 
  && \quad\text{if } e = e_1\ \lstinline+instanceof MyClass+~~~~~
    \label{eq:tr:instanceof} \\
  & \mathit{lib}.m\lstinline+(exorcise(+e_1\lstinline+)+,\cdots\lstinline+)+ 
  && \quad\text{if } e = \label{eq:tr:externalcall} 
  	\mathit{lib}.m\lstinline+(+e_1,\cdots,e_k\lstinline+)+ \\
  & \lstinline+nullUnbox(+ \mathit{unbox}(e_1) \lstinline+)+ 
  && \quad\text{if } e = \mathit{unbox}(e_1)
    \label{eq:tr:unbox} \\
  & \lstinline+nullDeref(+ e_1 \lstinline+)+.f
  && \quad\text{if } e = e_1.f \label{eq:tr:deref} \\
  & e && \quad\text{otherwise} \nonumber
\end{align}
\end{subequations}
\lstset{basicstyle=\ttfamily\normalsize} 
\caption{\ourtool's expression transformations (excepting method calls) : rules 
  \ref{eq:tr:equalnull}--\ref{eq:tr:externalcall}
preserve the semantics of the program under debug (\autoref{sec:sp}); rules
\ref{eq:tr:unbox}--\ref{eq:tr:method:call} inject calls to collect the 
\nullunboxed, \nullderef and \nullparamcall 
causality links (\autoref{sec:nullcausalitytrace}); in \autoref{eq:tr:deref},
$f$ denotes either a function or a field.  These rules are applied simultaneously to those in \autoref{fig:tr:stmt}.}
\label{fig:tr:expr}
\end{figure*}

\subsubsection{Overview}

Our idea is to inject a number of method calls in the program under debug.
They are listed in Table \ref{tab:explanation-injected-methods}.
For instance, \texttt{o = foo()} is transformed into \texttt{o = nullAssign(foo(), ``o, line 24'')} where  method ``nullAssign'' logs whether x is null at this assignment, returns x if x is a valid object or a ghost, or a new ghost if x is null.
There are a number of other methods and transformations that we now explain.

\autoref{fig:tr:expr} and \autoref{fig:tr:stmt} define \ourtool's 
transformations.  We have separated these rules into these two figures for
clarity;  in practice, the rules in these two figures are
applied simultaneously to a program.  \autoref{fig:tr:expr} contains rewriting
rules for expressions, with the exception of method calls;
\autoref{fig:tr:stmt} contains rules for statements and includes method calls,
which are both expressions and statements.
 In the figures, $e$ and $e_n$ are Java expressions, and $s$ is a
statement.
 For brevity, \autoref{fig:tr:stmt} introduces
$\langle\mathit{method\_decl}\rangle$
 for a method declaration and 
$\langle\mathit{method\_body}\rangle$ for its body.
  Since version 5.0, Java 
automatically boxes and unboxes
 primitives to and from object wrappers,
$\mathit{unbox}(e_1)$ denotes
 this operation.
Equations \ref{eq:tr:equalnull}--\ref{eq:tr:externalcall}, in
\autoref{fig:tr:expr}, define our semantics-preserving expression
transformations. \autoref{sec:sp} discusses them.

The equations inject the following functions into the
program under debug: \lstinline+nullDeref+, \lstinline+nullParam+, \lstinline+nullPassed+,  \lstinline+nullAssign+, and \lstinline+nullReturn+.
These functions all check their argument for nullity.  If their argument is
\nullv, they create a null ghost and add the causality link built into their
name, i.e. \lstinline+nullAssign+ adds \nullassign.  If their argument is a
null ghost, they all append the appropriate causality link to the causality trace.

\begin{figure*}[t]
\begin{subequations}
  \label{eq:tr:stmt}
\begin{align}[left = {T(s)=\empheqlbrace}]
  & o\ \leftarrow\ \lstinline+nullAssign(+ e \lstinline+);+\hspace{5.5cm}
  && \text{if } s = o\ \leftarrow\ \lstinline++ e \label{eq:tr:assignnullexpr}\hspace{2.6cm} \\
  & o\ \leftarrow\ \lstinline+nullAssign(null);+
  && \text{if } s = o \label{eq:tr:assignnulllit} \\
   & m\lstinline+(nullParam(+p_1, \cdots\lstinline+)+\lstinline+)+
  && \text{if } s = m\lstinline+(+p_1, \cdots,p_n\lstinline+)+ \label{eq:tr:method:call} \\
 & \begin{minipage}{5cm}
m($p_1$, \ldots, $p_n$) \{\\
~~$\forall p_i $, $p_i\leftarrow \lstinline+nullPassed+(p_i\lstinline+);+$\\
~~ $\langle\mathit{method\_body}\rangle$ \}\\
\end{minipage}
  && \text{if } s = \langle\mathit{method\_decl}\rangle \label{eq:tr:nullparamentry} \\
  & \lstinline+return nullReturn(+ e \lstinline+);+ 
    && \text{if } s = \lstinline+return+\ e \lstinline+;+ 
    \label{eq:tr:stmt:return} \\
  & s && \text{otherwise} \nonumber
\end{align}
\end{subequations}
\caption{\ourtool's statement transformations (including method calls) : these 
rules inject calls into statements to collect the \nullliteral, \nullassign,
\nullparamentry, and \nullreturn causality links
(\autoref{sec:nullcausalitytrace});
$s$ denotes a statement;
$\langle\mathit{method\_decl}\rangle$
denotes a method declaration and 
$\langle\mathit{method\_body}\rangle$ its body; 
 and $p_i$ binds to a function's formals in a declaration
and actuals in a call.  These rules are applied simultaneously to those in
\autoref{fig:tr:expr}.}
\label{fig:tr:stmt}
\end{figure*}

\subsubsection{Detection of \nullvs}

To provide the origin and causality of null dereferences, one has to detect
null values \emph{before} they become harmful, i.e. before they are
dereferenced.  This section describes how \ourtool's transformations inject
helper methods that detect and capture \nullvs.

In Java, as with all languages that prevent pointer arithmetic, \nullvs originate in
explicit or implicit literals within the application under debug or from an external library.  \nullliteral in
\autoref{table:causalitylinks} lists four examples of the former case.  \ourtool
statically detects \nullvs appearing in each of these contexts.  For explicit
\nullv assignment, as in \lstinline+x = null+, it applies its
\autoref{eq:tr:assignnullexpr}; for implicit, \lstinline+Object o;+, it applies
\autoref{eq:tr:assignnulllit}.  Both of these rewritings inject
\lstinline+nullAssign+, which instantiates a null ghost and starts its causality
trace with \nullliteral--\nullassign.  \autoref{eq:tr:stmt:return} injects
\lstinline+nullReturn+ 
to handle \lstinline+return null+, collecting \nullreturn.

Not all \nullvs can be statically detected: a library can produce them.  
An application may be infected by a \nullv from an external library in four 
cases:
\begin{inparaenum} 
\item assignments whose right hand side involves an external call;
\item method invocations one of whose parameter expressions involves an
  external call; 
\item boolean or arithmetic expressions involving external calls; and
\item callbacks from an external library into the program under debug.
\end{inparaenum}

\autoref{eq:tr:assignnullexpr} handles
the assignment case, injecting the \lstinline+nullAssign+ method to collect 
the causality links \nullreturn, \nullassign.
\autoref{eq:tr:externalcall} wraps the parameters of internal method calls 
with \lstinline+nullParam+, which handles external calls in a parameter list
and adds the \nullreturn and \nullparamentry links to a null ghost. 
\autoref{eq:tr:unbox} handles boolean or arithmetic expressions.
It injects the \lstinline+nullUnbox+ method to check nullity and create a null ghost or 
update an
existing one's trace with \nullreturn and \nullunboxed.  
Finally, \autoref{eq:tr:nullparamentry} handles library callbacks.  A 
library call back 
 happens when the application under debug provides 
an object
  to the library and the library invokes a method of this object, as in the
  ``Listener'' design pattern.  In this case, the library can bind
  \nullv to one of the method's parameters. 
  Because we cannot know which method may be involved in a callback,
  \autoref{eq:tr:nullparamentry}
  inserts a check for each argument at the beginning of every
  method call, potentially adding \nullparamentry to a ghost's causality trace.

Rewriting Java in the presence of its
\lstinline+final+ keyword is challenging.  
\autoref{fig:example-param-encaps} shows an example application of
\autoref{eq:tr:nullparamentry}. 
The first method is the application
 method and the second one is
the method after instruction by \ourtool.  The use
 of \lstinline+final+
variables, which can only be assigned once in Java,
 requires us to duplicate
the parameter as a local variable.  Renaming the
 parameters (b to b\_dup), then creating a
local variable with the same name as the original
 parameter, allows \ourtool
to avoid modifying the body of the method.

\begin{lstlisting}[float, numbers=none,captionpos=b,caption={Illustration of the Source Code Transformation for Causality Connection ``Null Method Parameter'' (\nullparamentry).},label=fig:example-param-encaps]
//initial method
void method(Object a, final Object b){
    //method body
}

//is transformed to
void method(Object a, final Object b_dup){
    a = NullDetector.nullPassed(a);
    b = NullDetector.nullPassed(b_dup);
    //method body
}
\end{lstlisting}

\subsection{Semantics Preservation}
\label{sec:sp}

Using null ghosts instead of \nullvs must not modify program execution.
\ourtool therefore defines the three transformations in Equations
\ref{eq:tr:equalnull}--\ref{eq:tr:externalcall}, whose aim is
to preserve semantics.  We evaluate the degree to which our transformations
preserve application semantics in \autoref{sec:eval}.

\paragraph{Comparison Operators} Consider ``\lstinline{o == null}".  When
\lstinline+o+ is \nullv, \lstinline+==+ evaluates to true.  If,
however, \lstinline{o} points to a null ghost, the expression evaluates to
false.  \autoref{eq:tr:equalnull} preserves the original behavior by rewriting expressions,
 to include the conjunct 
``\lstinline{!o instanceof NullGhost}". Our example ``\lstinline{o == null}"
becomes the expression ``\lstinline{o == null && !o instanceof NullGhost}".  
Here, \lstinline{NullGhost} is a marker interface that all null
  ghosts implement.  The rewritten expression is equivalent to the original,
  over all operands, notably including null ghosts. 

Java developers can write ``\lstinline{o instanceof MyClass}" to check the
compatibility of a variable and a type.  Under Java's semantics, if
\lstinline{o} is null, no error is thrown and the expression returns false.
When \lstinline{o} is a null ghost, however, the expression returns true.
\autoref{eq:tr:instanceof} solves this problem. To preserve behavior, it
rewrites appearances of the \lstinline+instanceof+ operator, e.g. replacing
``\lstinline{o instanceof MyClass}'' with 
``\lstinline{o instanceof MyClass && !o instanceof NullGhost}".  

\paragraph{Usage of Libraries}
During the execution of a program that uses libraries, one may pass
\nullv as a parameter to a library call.
For instance, \lstinline+o+ could be \nullv when \lstinline{lib.m(o)} executes.
After \ourtool's transformation, \lstinline+o+ may be bound to a null ghost.
In this case, if the library checks whether its parameters are null, 
using \lstinline{x == null} or \lstinline{x instanceof SomeClass}, a null ghost
could change the behavior of the library and consequently of the program.
Thus, for any method whose source we lack, we modify its calls to ``unbox the
null ghost'', using \autoref{eq:tr:externalcall}.
In our example, \lstinline{lib.m(o)} becomes \lstinline{lib.m(exorcise(o))}.
When passed a null ghost, the method \lstinline{exorcise} returns
the \nullv that the ghost wraps.

\paragraph{Emulating Null Dereferences} When dereferencing a \nullv,
Java throws an exception object  \lstinline+NullPointerException+.
When dereferencing a null ghost, the execution must also results in
throwing the same exception.
In \autoref{fig:example-ghost-class}, a null ghost 
throws the exception \lstinline+CasperNullPointerException+, which 
extends Java's exception \lstinline+NullPointerException+.
The \ourtool's specific exception contains the
dereferenced null ghost and overrides the usual exception reporting methods, 
namely the \lstinline+getCause+, \lstinline+toString+, and \lstinline+printStackTrace+ methods,
to display the ghost's causality trace.

Java throws a \lstinline+NullPointerException+ in three cases: 
\begin{inparaenum}[\itshape a\upshape)]
\item a method call on \nullv;
\item a field access on \nullv; or
\item unboxing a \nullv from a primitive type's object wrapper.
\end{inparaenum}
\ourtool trivially emulates method calls on a \nullv: it defines each method in
a ghost to throw \lstinline+CasperNullPointerException+, as
\autoref{fig:example-ghost-class} shows.  Java does not provide a listener 
that monitors field accesses.  \autoref{eq:tr:deref} overcomes 
this problem; it wraps
expressions involved in a field access in \lstinline+nullDeref+, which 
checks for a null ghost, prior to the field access.  For instance, \ourtool
transforms \lstinline+x.f+ into \lstinline+nullDeref(+$e$\lstinline+).f+.
Since version 5.0, Java has supported autoboxing and unboxing to facilitate
working with its primitive types.  A primitive type's object wrapper may
contain a \nullv;  if so, unboxing it triggers a null value triggers a null
dereference error.  For example, \lstinline+Integer a = null; int b = a+,
\lstinline*a + 3* or \lstinline+a * 3+ all throw
\lstinline+NullPointerException+.

\subsection{Implementation}
\label{sec:imp}

\ourtool requires, as input, the source code of the program under debug,
together with the binaries of the dependencies.  Its transformations are
automatic and produce an instrumented version of the program under debug.
We stack source code transformation at compile time and dynamic binary code transformation at load time.
The reasons are low-level details that are specific to the Java platform as explained above.

\paragraph{Source Code Transformations} We perform our source code transformations using Spoon~\cite{spoon}.
This is done at compile time, just before the compilation to bytecode.
Spoon performs all modifications on a model representing the AST of the program under debug.  Afterwards, Spoon
generate new Java files that contain the program corresponding to the AST after application of the transformations of \autoref{fig:tr:expr} and \autoref{fig:tr:stmt}. 

\paragraph{Binary Code Transformations} We create null ghosts with binary code
generation using ASM\footnote{\url{http://asm.ow2.org/}}. The reason is the Java \lstinline+final+ keyword.  This keyword can be
applied to both types and methods and prevents further extension.
Unfortunately,  we must be able override any arbitrary class and all methods to create
null ghost classes.  To overcome this protection at runtime, \ourtool uses its
our own classloader, which ignores the final keyword in method signatures when
the class is loaded.  For example, when \lstinline+MyClass+ must be
``haunted'', the class loader generates \lstinline+MyClassGhost.class+ on the fly. 

\paragraph{Limitations}\label{sec:limit} \ourtool cannot identify the root cause of a null
pointer dereference in two cases.  The first is when the root cause is in
external library code that we cannot rewrite. This is the price we pay to avoid
assuming a closed world, where all classes are known and manipulatable.  The
second is specific to the fact that we implemented \ourtool in Java: our
classloader technique for overriding \lstinline+final+ classes and methods does
not work for JDK classes, because most of these classes are loaded before the
invocation of application-specific class loaders.  One consequence is
that our implementation of \ourtool cannot provide causality traces for Java strings.

\section{Empirical Evaluation}
\label{sec:eval}

We now evaluate the capability of our approach to build correct causality traces of real errors from large-scale open-source projects.
The evaluation answers the following research questions:

\newcommand\rqA{$RQ1$: Does our approach provide the correct causality trace?\xspace}
\newcommand\rqB{$RQ2$: Do the code transformations preserve the semantics of the application?\xspace}
\newcommand\rqC{$RQ3$: Is the approach useful with respect to the fixing process?\xspace}

\noindent\rqA\\
\rqB\\
\rqC\\

RQ1 and RQ2 concern correctness. In the context of null deference
analysis, 
RQ1 focuses on one kind of correctness defined as the capability to provide the root cause
of the null dereference. In other words, the causality trace has to connect the
error to its root cause.
RQ2 assesses that the behavior of the application under study does not vary after applying our code transformations.
RQ3 studies the extent to which causality traces help a developer to fix null
dereference bugs.

\begin{table*}
\begin{tabularx}{\textwidth}{p{2.5cm}|X|X}
\textbf{ \# Bug Id} & \textbf{Problem summary} & \textbf{Fix summary} \\
\hline
 McKoi & new JDBCDatabaseInterface  with null param -> field -> deref& Not fixed (Artificial bug by  \cite{bond2007tracking})\\
\hline
 Freemarker \#107 &circular initialization makes a field null in WrappingTemplateModel -> deref & not manually fixed. could be fixed manually by adding hard-code value. no longer a problem with java 7. \\
\hline
 JFreeChart \#687 & no axis given while creating a plot & can no longer create a plot without axis modifying constructor for fast failure with error message \\
\hline
 collection \#331 & no error message set in a thrown NPE & add a check not null before the throw + manual throwing of NPE\\
\hline
 math \#290 & NPE instead of a domain exception when a null List provided& normalize the list to use empty list instead of null\\
\hline
 math \#305 & bad type usage (int instead of double). Math.sqrt() call on an negative int -> return null. should be a positive double & change the type\\
\hline
math \#1117 & Object created with too small values, after multiple iterations of a call on this object, it returns null & create a default object to replace the wrong valued one \\
\hline
7 other bugs &  & add a nullity check  \\
\end{tabularx}
\caption{A dataset of 14 real null dereference errors from large scale open-source projects. The dataset is made publicly available for future replication and research on this problem.}
\label{fig:tab-dataset}
\end{table*}

\begin{table}
\centering\begin{tabular}{p{2.5cm}|r|r}
Bug ID & \#LOC & \#classes \\
\hline
McKoi 	 & 48k & 275 \\
Freemarker \#107& 37k & 235 \\
JFreeChart \#687& 70k & 476 \\
coll-331		& 21k & 256 \\
lang-304	& 17k & 77 \\
lang-587	& 17k & 80 \\
lang-703	& 19k & 99 \\
math-290	& 38k & 388 \\
math-305	& 39k & 393 \\
math-369	& 41k & 414 \\
math-988a	& 82k & 781 \\
math-988b	& 82k & 781 \\
math-1115	& 90k & 885 \\
math-1117	& 90k & 885 \\
\end{tabular}
\caption{Descriptive Summary of the Dataset of Null Dereferences}
\label{tab:desc-stats}
\end{table}

\subsection{Dataset}

We built a dataset of real life null dereference bugs.
There are two inclusion criteria.
First, the bug must be a real bug reported on a publicly-available forum (e.g. a bug repository).
Second, the bug must be reproducible.

The reproducibility is challenging. Since our approach is dynamic, we must be able to compile and run the software in its faulty version.
First, we need the source code of the software at the corresponding buggy version.
Second, we must be able to compile the software.
Third, we need to be able to run the buggy case.
In general, it is really hard to reproduce real bugs in general and null dereferences in particular.
Often, the actual input data or input sequence triggering the null dereference is not given, or the exact buggy
version is not specified, or the buggy version can no longer be compiled and executed.

We formed our data set in two ways.  First, we tried to replicate results over a published data set ~\cite{bond2007tracking} as described below.  Second, we selected a set of popular projects.  For each project, we used a bag of words over their bug repository (e.g. bugzilla of jira) to identify an under approximate set of NPEs.  We then faced the difficult challenge of reproducing these bugs, as bug reports rarely specify the bug-triggering inputs.  Our final data set is therefore conditioned on reproducibility.  We do not, however, have any reason to believe that any bias that may exist in our data set would impact Casper general applicability.

Under these constraints, we want to assess how our approach compares to the closest related work~\cite{bond2007tracking}.
Their dataset dates back to 2008.
In terms of bug reproduction 6 years later, this dataset is hard to replicate.
For 3 of the 12 bugs in this work, we cannot find any description or bug report.
For 4 of the remaining 9, we cannot build the software because the versions of the libraries are not given or no longer available.
3 of the remaining 5 do not give the error-triggering inputs, or they do not produce an error.
Consequently, we were only able to reproduce 3 null dereference bugs from Bond et al.'s dataset.

We collected 7 other bugs. The collection methodology follows.
First, we look for bugs in the Apache Commons set of libraries (e.g. Apache Commons Lang).
The reasons are the following.
First, it is a well-known and well-used set of libraries.
Second, Apache commons bug repositories are public, easy to access and search.
Finally, thanks to the strong software engineering discipline of the Apache foundation, a failing test case is often provided in the bug report.

To select the real bugs to be added to our dataset we proceed as follows.
We took all the bugs from the Apache bug repository\footnote{\url{https://issues.apache.org/jira/issues}}.
We then select 3 projects that are well used and well known (Collections, Lang and Math).
We add the condition that those bug reports must have ``NullPointerException" (or ``NPE") in their title.
Then we filter them to keep only those which have been fixed and which are closed (our experimentation needs the patch).
After filters 19 bug reports remain\footnote{The link to automatically set those filters is given in \projecturl}.
Sadly, on those 19 bug reports, 8 are not relevant for our experiment:
3 are too olds and no commit is attached (COLL-4, LANG-42 and Lang-144),
2 concern Javadoc (COLL-516 and MATH-466),
2 of them are not bugs at all (LANG-87 and MATH-467),
1 concerns a VM problem.
Finally, we add the 11 remaining cases to our dataset.

Consequently, the dataset contains the 3 cases from \cite{bond2007tracking} (Mckoi, freemarker and jfreechart) and 11 cases from Apache Commons (1 from collections, 3 from lang and 7 from math).
In total, the bugs come from 6 different projects, which is good for assessing the external validity of our evaluation.
This makes a total of 14 real life null dereferences bugs in the dataset.

Table~\ref{fig:tab-dataset} shows the name of the applications, the number of the bug Id (if existing), a summary of the NPE cause and a summary of the chosen fix.
We put only one line for 7 of them because they use the same simple fix (i.e. adding a check not null before the faulty line).
The application coverage of the test suites under study are greater than 90\% for the 3 Apache common projects (11 out of the 14 cases).
For the 3 cases from \cite{bond2007tracking} (Mckoi, freemarker and jfreechart), we do not have access to the full test suites.
Table \ref{tab:desc-stats} gives the main descriptive statistics.
For instance, the bug in McKoi is an application of 48000+ lines of code spread over 275 classes.

This dataset only contains real null dereference bugs and no artificial or toy bugs. To reassure the reader about cherry-picking, we have considered all null dereferenced bugs of the 3 selected projects. We have not rejected a single null dereference that \ourtool fails to handle.

\subsection{Methodology}
\subsubsection{Correctness}
\label{sec:correctness}

\emph{\rqA} To assess whether the provided element is responsible for a null dereference, we manually analyze each case.
We manually compare the result provided by our technique with those coming from a manual debugging process that is performed using the debug mode of Eclipse.

\emph{\rqB} To assert that our approach does not modify the behavior of the application, we use two different strategies.

First, we require that the origin program and the transformed program both pass and fail the 
same tests in the test suite (when it exists).  
Having the same behavior according to a test suite does not prove semantic equivalence, it is only an indication  the program has not been broken. 
However, this is the only possible technique since current research on formal equivalence is not capable of proving semantic equivalence in a programming language as complex as Java.

This test suite test only addresses the correctness of externally observable behavior of the program under debug.
To assess that our approach does not modify the internal behavior, we
compare the execution traces of the original program
(prior to code transformation) and the program after transformation.
Here, an ``execution trace'' is the ordered list of all method calls and of all returned values, executing over the entire test suite.
This trace is obtained by logging method entry (injecting 
\lstinline+package.class\#method( arg.toString \ldots)+) and 
logging return (injecting \lstinline+package.class\#method():returnedValue.toString+).
We filter out all calls to the Casper framework, then align the two traces.  They must be identical.
As for the test suite, execution trace equivalence is only a proxy to complete equivalence.

\subsubsection{Effectiveness}
\emph{\rqC}
To assert that our additional data is useful, we look at whether the location of the real fix is given in the causality trace.
If the location of the actual fix is provided in the causality trace, it would have helped the developer by reducing the search space of possible solutions.
Note  that in 7/14 cases of our dataset, the fix location already appears in the original stack trace.
Those are the 7 simple cases where a check not null is sufficient to prevent the error.
Those cases are valuable in the context of our evaluation to check that:
1) the variable name is given (as opposed to only the line number of a standard stack trace),
2) the causality trace is correct (although the fix location appears in the  original stack trace, it does not prevent a real causality trace with several causal connections).

\begin{table*}
\centering\begin{tabularx}{17cm}{X|X|X|X|X|X|X}
\# Bug Id & Fix location & \scriptsize{Fix location in the standard stack trace} & Addressed by \cite{bond2007tracking} & \scriptsize{Fix location in \ourtool's  causality trace} & Causality Trace &  Exec. time instrumented (ms) \\ 
\hline
 McKoi & Not fixed & No & No & Yes & \nullliteral-\nullassign-\nullreturn-\nullassign-\nullderef &  382\\
\hline
 Freemarker \#107& Not fixed & No & Yes & Yes  & \nullliteral-\nullassign-\nullderef &  691\\
\hline
 JFreeChart \#687 & FastScatterPlot 178 & No & No & Yes & \nullliteral-\nullparamentry-\nullparamcall-\nullderef &  222\\
\hline
 collection \#331 & CollatingIterator 350 & No & No  & Yes &\nullliteral-\nullassign-\nullderef&81 \\
\hline
 math \#290 & SimplexTableau 106/125/197 & No & No & No&\nullderef&107 \\
\hline
 math \#305 & MathUtils 1624 & No & No & No & \nullliteral-\nullreturn-\nullassign-\nullreturn-\nullderef& 68 \\
\hline
math \#1117 & PolygonSet 230 & No & No & No & \nullliteral-\nullassign-\nullreturn-\nullassign-\nullderef & 191\\
\hline
7 simple cases &  & Yes & Yes & Yes & \nullliteral-\nullassign-\nullderef  (x6) \newline \nullliteral-\nullassign-\nullunboxed &147 (average)\\
\hline
\hline
Total & & 7 / 14  & 8/14 & 11/14 & & \\ 
\end{tabularx}
\caption{Evaluation of \ourtool: in 13/14 cases, a causality trace is given, in 11/13 the causality trace contains the location where the actual fix was made.}
\label{fig:tab-result}
\end{table*}

\subsection{Results}

\subsubsection{RQ1} After verification by manual debugging, in all the cases under study, the element identified by our approach is the one responsible for the error. This result can be replicated since our dataset and our prototype software are made publicly available.

\subsubsection{RQ2} All the test suites have the same external behavior with and without our modifications according to our two behavior preservation criteria.
First, the test suite after transformation still passes.
Second, for each run of the test suite, the order of method calls is the same and the return values are the same.
In short, our massive code transformations do not modify the behavior of the program under study and provide the actual causality relationships.

\subsubsection{RQ3}
We now perform two comparisons.
First, we look at whether the fix locations appear in the standard stack traces.
Second, we compare the standard stack trace and causality trace to see whether the additional information corresponds to the fix.

Table~\ref{fig:tab-result} presents the fix locations (class and line number) (second column) and whether:
this location is provided in the basic stack trace (third column);
2) the location is provided by previous work \cite{bond2007tracking}  (fourth column);
3)  it is in the causality trace (last column). 
The first column, ``\# Bug Id'', gives the id of the bug in the bug tracker of the project (same as Table \ref{fig:tab-dataset}).

In 7 out of 14 cases (the 7 simple cases), the fix location is in the original stack trace.
For those 7 cases, the causality trace is correct and also points to the fix location.
In comparison to the original stack trace, it provides the name of the root cause variable.

In the remaining 7 cases, the fix location is not in the original stack trace.
This means that in 50\% of our cases, there is indeed a cause/effect chasm, that is hard to debug \cite{eisenstadt1997my}, because no root cause information is provided to the developer by the error message.
We now explain in more details those 7 interesting cases.

The related work  \cite{bond2007tracking} would provide the root cause in only 1 out of those 7 cases (according to an analysis, since their implementation is not executable).
In comparison, our approach provides the root cause in 4 out of those 7 cases.
This supports the claim that our approach is able to better help the developers in pinpointing the root cause compared to the basic stack trace or the related work.

\subsubsection{Detailed Analysis}

\paragraph{Case Studies}
There are two different reasons why our approach does not provide the fix location:
First, for one case, our approach is not able to provide a causality trace.
Second, for two cases, the root cause of the null dereference is not the root cause of the bug.

In the case of Math \#290, our approach is not able to provide the causality trace.
This happens because the null value is stored in an Integer, which is a final type coming from the jdk.
Indeed, \lstinline+java.lang.Integer+ is a native Java type and our approach cannot modify them (see \autoref{sec:limit}).

In the case of Math \#305, the root cause of the null dereference is not the root cause of the bug.
The root cause of this null dereference is shown in Listing~\ref{fig:math-305-src}.
The null responsible of the null dereference is initialized on line 4, the method call \texttt{distanceFrom} on line 6 return \texttt{NaN}, due to this NaN, the condition on line 7 fails, and the null value is returned  (line 9).
Here, the cause of the dereference is that a null value is returned by this method.
However, this is the root cause of the \texttt{null} but this is not the root cause of the bug.
The root cause of the bug is the root cause of the \texttt{NaN}.
Indeed, according to the explanation and the fix given by the developer the call \texttt{point.distanceFrom(c.getCenter())} should not return \texttt{NaN}.
Hence, the fix of this bug is in the \texttt{distanceFrom} method, which does not appear in our causality chain because no \texttt{null} is involved.

\begin{lstlisting}[float, basicstyle=\scriptsize, numbers=left,captionpos=b,caption={An excerpt of Math \#305 where the causality trace does not contain the fix location.},label=fig:math-305-src]

    private static Cluster<T> getNearestCluster(final Collection<Cluster> clusters, final T point) {
        double minDistance = Double.MAX_VALUE;
        Cluster<T> minCluster = null; //initialisation
        for (final Cluster<T> c : clusters) {
            final double distance = point.distanceFrom(c.getCenter()); //return NaN
            if (distance < minDistance) { //failing condition
                minDistance = distance;
                minCluster = c;
            }
        }
        return minCluster; //return null
    }
\end{lstlisting}

In the case of Math \#1117,  the root cause of the null dereference is not the root cause of the bug.
The root cause of this null dereference is shown in Listing~\ref{fig:math-1117-src}.
The null responsible of the dereference is the one passed as second parameter of the constructor call on line 10.
This null value is stored in the field \lstinline+minus+ of this SplitSubHyperplane.
Here, the cause of the dereference is that a null value is set in a field of the object returned by this method.
Once again, this is the root cause of the \texttt{null} but this is not the root cause of the bug.
The root cause of the bug is the root cause of the failing condition \lstinline+global < -1.0e-10+.
Indeed, according to the explanation and the fix given by the developer the Hyperplane passed as method parameter should not exist if its two lines are too close from each other.
Here, this Hyperplane comes from a field of a PolygonSet.
On the constructor of this PolygonSet they pass a null value as a parameter instead of this ``irregular'' object.
To do that, they add a condition based on a previously existing parameter called \texttt{tolerance}, if the distance of the two lines are lower than this tolerance, it returns a null value. (It is interesting that the fix of a null dereference is to return a null value elsewhere.)

\begin{lstlisting}[float, basicstyle=\scriptsize, numbers=left,captionpos=b,caption={An excerpt of Math \#1117 where the causality trace does not contain the fix location.},label=fig:math-1117-src]

    public SplitSubHyperplane split(Hyperplane hyperplane) {
        Line thisLine  = (Line) getHyperplane();
        Line otherLine = (Line) hyperplane;
        Vector2D crossing = thisLine.intersection(otherLine);
 	   if (crossing == null) { // the lines are parallel
            double global = otherLine.getOffset(thisLine);
            return (global < -1.0e-10) ?
                   new SplitSubHyperplane(null, this) :
                   new SplitSubHyperplane(this, null); // initialisation
        }
	   ...
    }
\end{lstlisting}

\paragraph{Size of the Traces}
There are mainly two kind of traces encountered in our experiment.
First, the one of size 3 and of kind \nullliteral-\nullassign-\nullderef type.
The 7 obvious cases (were the fix location is in the stack trace) contains 6 traces of this kind.
In all those cases encountered in our experiment, the null literal has been assigned to a field.
This means that a field has not been initialized (or initialized to null) during the instance creation, hence, this field is dereferenced latter.
This kind of trace is pretty short so one may think that this case is obvious.
However, all of those fields are initialized long ago the dereference.
In other words, when the dereference occurs, the stack has changed and no longer contains the information of the initialization location.

Second, the one of size $\leq 4$ where the null is stored in a variable then passed as argument in one or multiple methods.
In all those case, the null value is either returned by a method at least once or passed as a parameter.

\paragraph{Execution Time}
\label{sec:overhead}

To debug a null dereference error, \ourtool  requires to instrument the code and to run the instrumented version. In all the cases, the instrumentation time is less 30 seconds. At runtime, \ourtool finds the causility trace of the failing input in less than 1 second (last column of Table~\ref{fig:tab-result}).
This seems reasonable from the developer viewpoint: she obtain  the causality trace in less than 30 seconds. We have also measured the overhead with respect the original test case trigerring the error: it's a 7x increase.

\section{Discussion}

\subsection{Causality Trace and Patch}

Once the causality trace of a null dereference is known, the developer can fix the dereference.
There are two basic dimensions for fixing null references based on the causality trace.

First, the developer has to select in the causal elements, the location where the fix should be applied: it is often at the root cause, i.e. the first element in the causality trace. It may be more appropriate to patch the code elsewhere, in the middle of the propagation between the first occurrence of the null and the dereference error.

Second, the developer has to decide whether there should be an object instead of null or whether the code should be able to gracefully handle the null value.
In the first case, the developer fixes the bug by providing an appropriate object instead of the null value.
In the second case, she adds a null check in the program to allow a null value.

Sometimes, the fix for a null dereference will be an insertion of a missing statement or the correction of an existing conditional. 
In this case, the \ourtool causality trace does not contain the exact location of the fix. However, as in the case of bug Math \#290 that we have discussed, it is likely that those modifications will be made in the methods involved in the causality trace. According to our experience, those cases are uncommon but future work is required to validate this assumption.

\subsection{Use in Production}
As shown in Section \ref{sec:overhead}, the overhead of the current implementation is too large to be used in production. 
We are confident that advanced optimization can reduce this overhead.
This is left to future work.

\subsection{Threats to Validity}
The internal validity of our approach and implementation has been assessed through RQ1 and RQ2: the causality trace of the 14 analyzed errors is correct after manual analysis. 
The threat to the external validity lies in the dataset composition:
does the dataset reflect the complexity of null dereference errors in the field? 
To address this threat, we took a special care in designing the methodology to build the dataset. It ensures that the considered bugs apply to large scale software and are annoying  enough to be reported and commented in a bug repository. 
The generalizability of our results to null dereference errors in other runtime environments (e.g. .NET; Python) is an open question to be addressed by future work.

\section{Related Work}

There are several static techniques to find possible null dereference bugs.
Hovemeyer et al.~\cite{hovemeyer2005evaluating} use byte-code analysis to provide possible locations where null dereference may happen.
Sinha et al.~\cite{sinha2009fault} use source code path finding to find the locations where a bug may happen and apply the technique to localize Java null pointer exceptions symptom location.
Spoto \cite{spoto2011precise} devises an abstract interpretation dedicated to null dereferences \cite{spoto2011precise}.
Ayewah and Pugh \cite{ayewah2010null} discussed the problems of null dereference warnings that are false positives.
Compared to these works, our approach is  dynamic and instead of predicting potential future bugs that may never happen in production, it gives the root cause of actual ones for which the developer has to find a fix.

Dobolyi and Weimer~\cite{dobolyi2008changing} present a technique to tolerate null dereferences based on the insertion of well-typed default values to replace the null value which is going to be dereferenced.
Kent~\cite{kent2008dynamic} goes further and, proposes two other ways to tolerate a null dereference: skipping the failing instruction or returning a well-typed object to the caller of the method.
In the opposite, our work is not on tolerating runtime null dereference but on giving advanced debugging information to the developer to find a patch.

The idea of identifying the root cause in a cause effect chain has been explored by Zeller~\cite{zeller2002isolating}. In this paper, he compares the memory graph from the execution of two versions of a same program (one faulty and one not faulty) to extract the instructions and the memory values which differ and presumably had lead to the error.
This idea has bee further extended by Sumner and colleagues \cite{sumner2009algorithms,sumner2013comparative}.
Our problem statement is different, those approaches takes as input two different versions of the program or two different runs and compare them.
On the contrary, we build the causality trace from a single execution.

The Linux kernel employs special values, called poison pointers, to transform certain latent null errors into fail-fast errors \cite{rubini2001linux}.
They share with null ghosts the idea of injecting special values into the execution stream to aid debugging and, by failing fast, to reduce the width of the cause/effect chasm.
However, poison values only provide fail-fast behavior and do not provide causality traces or even a causal relationship as we do.

Romano et al. \cite{romano2011approach} find possible locations of null dereferences by running a genetic algorithm to exercise the software. If one is found, a test case that demonstrate the null dereference is provided.
Their technique does not ensure that the null dereferences found are realistic and represent production problem. On the contrary, we tackle null dereferences for which the programmer has to find a fix. 

Wang et al. \cite{wang2013generating} describe an approach to debug memory errors in C code. What they call ``value propagation chain'' corresponds to our causality traces. They don't provide a taxonomy of causal elements as we do in \autoref{table:causalitylinks} and they take a great care of pointer arithmetic, which is irrelevant in our case.. Their transformations are at the level of x86 code using dynamic instrumentation, while we work on Java source code. This makes a major difference: all the transformations we have described are novel, and cannot be inferred or derived from Wang et al.'s work.

Bond et al.~\cite{bond2007tracking} present an approach to dynamically provide information about the root cause of a null dereference (i.e. the line of the first null assignment). The key difference is that we provide the complete causality trace of the error and not only the first element of the causality trace. As discussed in the evaluation (\autoref{sec:eval}), the actual fix of many null dereference bugs is not necessary done at the root cause, but somewhere up in the causality trace.  

Like null ghosts, the null object pattern replaces nulls with objects whose
interface matches that of the null-bound variable’s
type.  Unlike null ghosts, the methods of an instance
of the null object pattern are empty.  Essentially, the null object pattern
turns method NPEs into NOPs. To this extent, the refactoring proposed by \cite{gaitani2015automated}, does not help to debug null dereferences but avoids some of them.
In contrast, null ghosts collect null dereference causality traces that allow a
developer to localize and resolve an NPE.

\section{Conclusion}

In this paper, we have presented \ourtool, a novel approach for debugging null dereference errors.
The key idea of our technique is to inject special values, called ``null ghosts'' into the execution stream to aid debugging. 
The null ghosts collect the history of the null value propagation between its first detection and the problematic dereference, we call this history the `causality trace''.
We define \nbtransfo code transformations responsible for 1) detecting null values at runtime, 2) collect causal relations and enrich the causality traces; 3) preserve the execution semantics when null ghosts flow during program execution.
The evaluation of our technique on 14 real-world null dereference bugs from large-scale open-source projects shows that \ourtool is able to provide a valuable  causality trace.
Our future work consists in further exploring the idea of ``ghost'' for debugging other kinds of runtime errors such as arithmetic overflows.

\bibliographystyle{abbrv}
\bibliography{biblio}

\end {document}